\documentclass[12pt,onecolumn]{IEEEtran}

\usepackage{amsmath,amsfonts,amsthm,amssymb}
\usepackage{latexsym}
\usepackage{textcomp}
\usepackage{stfloats}
\usepackage{verbatim}
\usepackage{graphicx}
\usepackage{epstopdf}
\usepackage{float}
\usepackage{caption}
\usepackage{balance}
\usepackage{bm}
\usepackage{subfigure}
\usepackage{cite}
\usepackage[ruled]{algorithm2e}
\usepackage{xcolor}

\def\BibTeX{{\rm B\kern-.05em{\sc i\kern-.025em b}\kern-.08em
		T\kern-.1667em\lower.7ex\hbox{E}\kern-.125emX}}

\allowdisplaybreaks[4]

\newcommand{\rev}{}
\newcommand{\revp}{}

\begin{document}
\title{Channel Tracking for RIS-aided mmWave Communications Under High Mobility Scenarios}
\author{Yu Liu, Ming Chen, Cunhua Pan, Yijin Pan, Yinlu Wang, Yaoming Huang, Tianyang Cao and Jiangzhou Wang \IEEEmembership{Fellow, IEEE}
\thanks{(corresponding authors: Ming Chen, Cunhua Pan) \\
Y. Liu, M. Chen, C. Pan, Y. Pan, and Y. Wang  are with the National Mobile Communications Research Laboratory, Southeast University, Nanjing, 210096, China.
(e-mail: \{liuyu\_1994, chenming, cpan, panyj, yinluwang \}@seu.edu.cn).}
\thanks{Y.Huang and T.Cao are with China Mobile Group Design Institute Co., Ltd., China. (e-mail: \{caotianyang, huangyaoming\}@cmdi.chinamobile.com)}
\thanks{J. Wang is the School of Engineering and Digital Arts, University of Kent, Canterbury, United Kingdom. 
(e-mail:j.z.wang@kent.ac.uk).}
}
\maketitle

\begin{abstract}
The emerging reconfigurable intelligent surface (RIS) technology is promising for applications in the millimeter wave (mmWave) communication systems to effectively compensate for propagation loss or tackle the blockage issue. 
Considering the high mobility of users in realistic scenarios, it is essential to adjust the phase shifts in real time to align the beam towards the mobile users, which requires to frequently estimate the channel state information.
Hence, it is imperative to design efficient channel tracking schemes to avoid the complex channel estimation procedure.
In this paper, we develop a novel channel tracking scheme with two advantages over conventional schemes.
First, our tracking scheme is based on the cascaded angles at the RIS instead of the accurate angle values, which is more practical.
Second, it can be employed under a more general setting where the noise can be non-Gaussian.
Simulation results show the high tracking accuracy of our proposed scheme, and validate the superiority to the existing EKF-based tracking scheme.
\end{abstract}


\section{Introduction}
Communications at millimeter wave (mmWave) frequencies have attracted extensive research attention due to its advantages in providing high data rate transmission. 
From the electromagnetic propagation theory, the diffraction effect of radio signals weakens as the frequency increases \cite{mmwavechannel}. 
Hence, mmWave communication relies heavily on the existence of line-of-sight (LoS) path to maintain a satisfactory communication quality. 
Unfortunately, in mmWave mobile communication scenarios, the path between the transmitter and the receiver is usually blocked by obstacles, which severely degrades the system performance. 
Recently, the reconfigurable intelligent surfce (RIS) has become promising in addressing the blockage issue in an energy-efficient and cost-effective way.
An RIS is a planar surface composed of a large number of low-cost passive reflecting elements, which can adjust the incident signals via inducing the additional phase shifts \cite{CPan, survey2}. 
The RIS can be deployed to establish a virtual-line-of-sight (VLoS) path to bypass the blockage in wireless environment, which avoids communication outage when the user moves into blind area.

Moreover, with efficient phase shifts design, an RIS can reflect the signals in a desired direction, which significantly enhances the communication quality \cite{2021RISwangrui}. 
To the best of our knowledge, most of the existing contributions on beam alignment in RIS-aided wireless systems focused on the stationary scenario, where the positions of the base station, the RIS and the user remain almost stable. 
In stationary scenarios, it is reasonable to acquire the channel state information (CSI) via channel estimation algorithms.
However, under the high mobility scenarios,  the computational complexity and time consumption of such estimation procedure is extremely high, which leads to the frequent communication outage for high-mobility users.
Hence, to avoid periodically complicated channel estimation, the filter-based channel tracking schemes were proposed, which are the online CSI update approaches based on the initial estimation and the subsequently received signals \cite{Filter_tutorial}.

In conventional communication scenarios without RISs, filter-based tracking schemes are commonly based on the angle-of-departure (AoD) and angle-of-arrival (AoA) obtained via the initial estimation \cite{EKF1,EKF2,PF1}.
However, in the RIS-aided communication system, it is intractable to acquire the AoD and AoA at the RIS since RIS does not possess active radio frequency (RF) chains.
\revp{
The authors in \cite{Dai} proposed a novel two-timescale channel estimation framework to estimate the parts of the cascaded channel separately. 
}
Recently, the authors of \cite{zhougui} proposed a novel channel estimation approach without acquiring the real angle information at the RIS. 
Instead of estimating the AoA and AoD, the authors estimated the cascaded angles at the RIS, which is more practical considering the passive property of RISs.
However, this approach is not applicable for the mobile scenarios since the fast variation of angles caused by user mobility makes the pilot overhead unaffordable.

Against the above background, for the RIS-aided mmWave mobile communication system, we propose a novel tracking scheme named particle filter with cascaded angles (PF-WCA) scheme. 
This scheme is based on the widely used particle filters \cite{Filter_tutorial}, and it chooses to track cascaded angles instead of AoA/AoD for practical considerations.
In addition, under the high mobility scenarios, the noise of the cascaded angles is commonly non-Gaussian \cite{wutuo}, and the widely used Kalman filter-based tracking schemes are not suitable since they depend on the Gaussian assumptions \cite{EKF_book}.
Our PF-WCA scheme solves this issue through deriving an approximation of the importance density for the particle filter.
The contributions of this paper are summarized as follows:
\begin{enumerate}
	\item We develop a novel channel tracking scheme named PF-WCA for the RIS-aided mobile mmWave system, which works well under the general assumption where the noise can be non-Gaussian in highly mobile environments.
	\item Simulation results validate the superiority of our PF-WCA scheme to the existing Kalman filter-based tracking schemes under high mobility scenarios, and reveal the impact of various design parameters on the tracking performance.
\end{enumerate}

\textit{Notations}: Matrices and vectors are denoted by bold uppercase letters and bold lowercase letters, respectively. $*$ and $H$ denote the conjugate and Hermitian transpose, respectively. $\otimes$ and $\odot$ denote the Kronecker product and Hadamard product, respectively.

\section{System Model}
We consider a narrow-band mmWave system, where a single-antenna user equipment (UE) communicates with a base station (BS) equipped with an $Nr$-antenna uniform linear array (ULA). The LoS path between the user and the BS is assumed to be blocked. Hence, a uniform planar array (UPA)-type RIS equipped with $L\times L$ passive reflecting elements is deployed to improve the communication performance. The BS aims to track the mobile user with the assistance of the RIS. 

\begin{figure}
	\centering
	\includegraphics[width=0.5\textwidth]{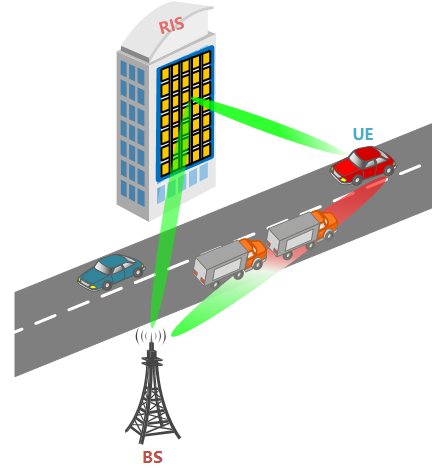}
	\caption{system model}
	\label{system model}
\end{figure}

To capture the impact of the mobility, we consider the channel variation model consisting of two timescales. We define the minimal time unit as a slot and $N$ successive slots as a block. It is assumed that the initial CSI can be acquired through the channel estimation at the beginning of each block, and the path loss of the channel remains invariant within the block. Between two successive slots, the beam angles in the RIS-BS link remain constant while the beam angles in the UE-RIS link vary due to the user mobility. 

The received signal at time slot $k$ can be modeled as 
\begin{equation}
\begin{aligned}
	y_k=& \alpha \sqrt{p}\mathbf{w}^H\left(\bar{\theta}\right)\bm{\alpha}_{RX}\left(\theta\right)\bm{\alpha}_{out}^H\left(\phi_{e},\phi_{a}\right)\bm{\Omega}_k \\
	& \bm{\alpha}_{in}\left(\psi_{e,k}, \psi_{a,k}\right)s_k+z_k,
\end{aligned}
\label{channel model1}
\end{equation}
where $s_k$ is the transmission signal with a known training symbol $1$, $z_k \sim \mathcal{CN}(0,\sigma^2)$ denotes the additive white Gaussian noise at time slot $k$, $\alpha$ is the channel gain employing the model in \cite{CPan}, $p$ is the transmission power (pTX), and $\bm{\Omega}_k=\mathrm{diag}\left(\mathbf{e}_k\right)$ denotes the phase shift matrix at time slot $k$ with the $L^2\times 1$ phase shift vector $\mathbf{e}_k$.
The combining vector $\mathbf{w}$ and array response vector $\bm{\alpha}_{RX}$ are separately given by
\begin{flalign}
    & \mathbf{w}\left(\bar{\theta}\right)=\frac{1}{\sqrt{N_r}}\left[1, e^{j\frac{2\pi d}{\lambda}\cos\bar{\theta}},\cdots,e^{j(N_r-1)\frac{2\pi d}{\lambda}\cos\bar{\theta}}\right]^T, \\
	& \bm{\alpha}_{RX}\left(\theta\right)=\frac{1}{\sqrt{N_r}}\left[1, e^{j\frac{2\pi d}{\lambda}\cos\theta},\cdots,e^{j(N_r-1)\frac{2\pi d}{\lambda}\cos\theta}\right]^T,
\end{flalign}
where $\lambda$ is the carrier wavelength, $d$ is the element spacing, $\bar{\theta}$ is the angle used for combining, and $\theta$ is the AOA at the BS. 
For the RIS, the $L^2\times1$ array response vector $\bm{\alpha}_{out}$ is expressed as
\begin{equation}
	\bm{\alpha}_{out}\left(\phi_e,\phi_a\right)=\bm{\alpha}_v\left(\varrho\right)\otimes\bm{\alpha}_h\left(\zeta\right),
\end{equation}
where $\phi_{a}$ is the azimuth AoD of the RIS-BS link, $\phi_{e}$ is the elevation AoD of the RIS-BS link, and
\begin{align}
	& \varrho=\cos\phi_e, \quad \zeta=\sin\phi_e\cos\phi_a ,\\
	& \bm{\alpha}_v\left(\varrho\right)=\left[1,e^{-j\frac{2\pi d}{\lambda}\varrho},\cdots,e^{-j(L-1)\frac{2\pi d}{\lambda}\varrho}\right]^T ,\\
	& \bm{\alpha}_h\left(\zeta\right) = \left[1,e^{-j\frac{2\pi d}{\lambda}\zeta},\cdots,e^{-j(L-1)\frac{2\pi d}{\lambda}\zeta}\right]^T .
\end{align}
The $L^2\times1$ array response vector $\bm{\alpha}_{in}$ is expressed as
\begin{equation}
\bm{\alpha}_{in}\left(\psi_{e,k},\psi_{a,k}\right)=\bm{\alpha}_v\left(\rho_k\right)\otimes\bm{\alpha}_h\left(\chi_k\right),
\end{equation}
where $\psi_{a,k}$ is the azimuth AoD of the RIS-BS link at time slot $k$, $\psi_{e,k}$ is the elevation AoD of the RIS-BS link at time slot $k$, and
\begin{align}
& \rho_k=\cos\psi_{e,k}, \quad \chi_k=\sin\psi_{e,k}\cos\psi_{a,k} ,\\
& \bm{\alpha}_v\left(\rho_k\right)=\left[1,e^{-j\frac{2\pi d}{\lambda}\rho_k},\cdots,e^{-j(L-1)\frac{2\pi d}{\lambda}\rho_k}\right]^T ,\\
& \bm{\alpha}_h\left(\chi_k\right) = \left[1,e^{-j\frac{2\pi d}{\lambda}\chi_k},\cdots,e^{-j(L-1)\frac{2\pi d}{\lambda}\chi_k}\right]^T .
\end{align}
Eq. (\ref{channel model1}) can be rewritten as 
\begin{equation}
	\begin{aligned}
		y_k&=\alpha \sqrt{p} \mathbf{w}^H\left(\bar{\theta}\right)\bm{\alpha}_{RX}\left(\theta\right)\bm{\alpha}_{out}^H\left(\phi_e,\phi_a\right) \\
		& \quad \quad \mathrm{Diag}(\bm{\alpha}_{in}(\psi_{e,k},\psi_{a,k}))\mathbf{e}_k s_k+z_k  \\
		&=\alpha \sqrt{p} \mathbf{w}^H\left(\bar{\theta}\right)\bm{\alpha}_{RX}\left(\theta\right)\bm{\alpha}_{RIS}^H\left(\phi_e,\phi_a,\psi_{e,k},\psi_{a,k}\right)\mathbf{e}_k s_k+z_k,
	\end{aligned}
	\label{rewrite channel model}
\end{equation}
where the cascaded response vector $\bm{\alpha}_{RIS}$ is given by 
\begin{equation}
	\bm{\alpha}_{RIS}\left(\phi_e,\phi_a,\psi_{e,k},\psi_{a,k}\right)=\bm{\alpha}_{out}\left(\phi_e,\phi_a\right)\odot\bm{\alpha}_{in}^{*}\left(\psi_{e,k},\psi_{a,k}\right).
\end{equation}

In \eqref{rewrite channel model}, only the elevation angle $\psi_{e,k}$ and azimuth angle $\psi_{a,k}$ change with time. Given $\psi_{e,k}$ and $\psi_{a,k}$ at time slot $k$, those angles at time slot $k+1$ are expressed as
\begin{flalign}
	& \psi_{e,k+1}=\psi_{e,k}+\xi_{k+1}, \\
	& \psi_{a,k+1}=\psi_{a,k}+\delta_{k+1},
\end{flalign}
where $\xi_k \sim \mathcal{U}(-\psi_s,\psi_s)$ and $\delta_k \sim \mathcal{U}(-\psi_r,\psi_r)$ are the angle variation caused by mobility, called process noise, and $\psi_s$ is the maximum amplitude of the variation at each time slot.
Due to the property of the passive RIS, it is intractable to obtain the accurate values of $\psi_{e,k}$, $\psi_{a,k}$, $\phi_{e}$, $\phi_{a}$ at the RIS. In practice, it is more tractable to obtain the cascaded angles $\cos\phi_e-\cos\psi_{e,k}$ and $\sin\phi_e\cos\phi_a-\sin\psi_{e,k}\cos\psi_{a,k}$ \cite{zhougui}. Denote the elevation component $\cos\phi_e-\cos\psi_e$ and the azimuth component $\sin\phi_e\cos\phi_a-\sin\psi_{e,k}\cos\psi_{a,k}$ by $x_e$ and $x_a$ separately, the hidden states needed to be tracked are formed as $\pmb{x}=\left\{x_e,x_a\right\}$.

\section{Proposed PF-WCA Tracking Scheme}
In this section, we introduce our PF-WCA tracking scheme to track the hidden states.  
In existing works, the extended Kalman filter (EKF) was widely employed in the mmWave channel tracking tasks \cite{EKF1, EKF2}. However, EKF is not capable of dealing with the case where the process noise is non-Gaussian \cite{EKF_book}. 
Therefore, we propose to employ the particle filter (PF) in our PF-WCA scheme to neutralize the effect of non-Gaussian process noise.

The PF generates $N_s$ random particles $\pmb{x}_{k}^{ip}$, where $ip$ is the particle index, and each particle represents one possible choice of the hidden states. Each particle has a normalized weight $\bar{\omega}_{k}^{ip}$, which satisfies
$
\sum_{ip=1}^{N_s}\bar{\omega}_{k}^{ip}=1.
$
The hidden states estimation $\hat{\pmb{x}}_k$ at time slot $k$ is approximated by the weighted sum of the particles as $\hat{\pmb{x}}_{k}=\sum_{ip=1}^{N_s}\bar{\omega}_{k}^{ip}\pmb{x}_{k}^{ip}$.

The particles $\pmb{x}_{k}^{ip}$ are drawn from the proposal importance density $q\left(\cdot\right)$. Commonly, $q\left(\cdot\right)$ is set as the posterior probability function $p\left(\pmb{x}_{k}|\pmb{x}_{k-1}^{ip}\right)$ \cite{PF1}. However, the hidden states in this paper include several cosine operations, which makes the posterior function intractable to be computed. Hence, we need to derive an approximation of the posterior function as the proposal density $q\left(\cdot\right)$. The detailed analysis is given in the following propositions.

\newtheorem{prop}{\bf Proposition}
\begin{prop}
From time slot $k$ to time slot $k+1$, the variation of elevation state $x_e$ can be approximated to follow the uniform distribution given by $ x_{e,k+1}\sim \mathcal{U}\left(x_{e,k}-\psi_s,x_{e,k}+\psi_s\right)$.
\end{prop}
\begin{IEEEproof}
From time slot $k$ to time slot $k+1$, the hidden state $x_e$ of the cascaded channel varies as follows:
\begin{equation}
	\begin{aligned}
		\!\! x_{e,k+1}&=\cos\phi_e-\cos\psi_{e,k+1} \\
		&=\cos\phi_e-\cos\left(\psi_{e,k}+\xi_{k+1}\right) \\
		&=\cos\phi_e-\cos\psi_{e,k}\cos\xi_{k+1}+\sin\psi_{e,k}\sin\xi_{k+1}.
	\end{aligned}
\end{equation}
Assuming that $\psi_s$ is relatively small, the variation can be approximated as
\begin{equation}
	\begin{aligned}
		x_{e,k+1}  & \approx \cos\phi_e-\cos\psi_{e,k}+\sin\psi_{e,k}\xi_{k+1} \\
		&=x_{e,k}+\sin\psi_{e,k}\xi_{k+1},
	\end{aligned}
\end{equation}
since $\sin \xi_{k+1} \approx \xi_{k+1}$ and $\cos \xi_{k+1}\approx 1$. 
Hence, the distribution of $x_{e,k+1}$ is approximated as $\mathcal{U}(x_{e,k}-|\sin\psi_{e,k}||\psi_s|, x_{e,k}+|\sin\psi_{e,k}||\psi_s|)$, which can be further approximated as $\mathcal{U}(x_{e,k}-\psi_s, x_{e,k}+\psi_s)$ since $|\sin\psi_{e,k}|\le 1$.
\end{IEEEproof}
\begin{prop}
From time slot $k$ to time slot $k+1$, the variation of the azimuth state $x_a$ can be approximated to follow the uniform distribution given by $x_{a,k+1}\sim\mathcal{U}\left(x_{a,k}-\psi_s-\psi_r-\psi_s\psi_r, x_{a,k}+\psi_s+\psi_r+\psi_s\psi_r\right)$.
\end{prop}
\begin{IEEEproof}
From time slot $k$ to time slot $k+1$, the hidden state $x_a$ of the cascaded channel varies as follows: 
\begin{equation}
	\begin{aligned}
		x_{a,k+1}&=\sin\phi_e\cos\phi_a-\sin\psi_{e,k+1}\cos\psi_{a,k+1} \\
		&=\sin\phi_e\cos\phi_a-\sin\left(\psi_{e,k}+\xi_{k+1}\right)\cos\left(\psi_{a,k}+\delta_{k+1}\right) \\
		&\begin{aligned}
			& = \sin\phi_e\cos\phi_a-\left(\sin\psi_{e,k}\cos\xi_{k+1}+\cos\psi_{e,k}\sin\xi_{k+1}\right) \\
			& \quad \left(\cos\psi_{a,k}\cos\delta_{k+1}-\sin\psi_{a,k}\sin\delta_{k+1}\right).
		\end{aligned}
	\end{aligned}
\end{equation}
Similarly, assuming $\psi_s$ and $\psi_r$ are relatively small enough, the variation can be approximated as
\begin{equation}
	\begin{aligned}
		x_{a,k+1} & \approx  \sin\phi_e\cos\phi_a-\left(\sin\psi_{e,k}+\cos\psi_{e,k}\xi_{k+1}\right)\cdot \\
		& \quad \left(\cos\psi_{a,k}-\sin\psi_{a,k}\delta_{k+1}\right) \\
		& = x_{a,k}-\cos\psi_{e,k}\cos\psi_{a,k}\xi_{k+1}+\sin\psi_{e,k}\sin\psi_{a,k}\delta_{k+1} \\
		& \quad + \cos\psi_{e,k}\sin\psi_{a,k}\xi_{k+1}\delta_{k+1},
	\end{aligned}
\end{equation}
Similar to Proposition 1, the distribution of $x_{a,k+1}$ can be approximated to follow the uniform distribution given by \\ $x_{a+1,k}\sim\mathcal{U}\left(x_{a,k}-\psi_s-\psi_r-\psi_s\psi_r, x_{a,k}+\psi_s+\psi_r+\psi_s\psi_r\right)$. 
\end{IEEEproof}
\revp{
\begin{prop}
The particle filter state estimate will converge to the true estimate as the number of particles tends to infinity.
\end{prop}
\begin{IEEEproof}
The detailed proof is referred to Appendix B.
\end{IEEEproof}
}

According to Proposition 1 and Proposition 2, the importance density $q\left(\cdot\right)$ is given by
\begin{equation}
	q\left(\pmb{x}_{k+1}|\pmb{x}_{k}^{ip}\right)\sim\begin{Bmatrix}
	\mathcal{U}\left(x_{e,k}-\psi_s,x_{e,k}+\psi_s\right) \\
	\mathcal{U}(x_{a,k}-\psi_s-\psi_r-\psi_s\psi_r,  \\
	\quad x_{a,k}+\psi_s+\psi_r+\psi_s\psi_r)
\end{Bmatrix}. 
\end{equation}

The weight $\omega_k^{ip}$ associated with the particles can be computed as
\begin{equation}
	\omega_{k}^{ip} = p(y_k|\pmb{x}_{k}^{ip}) = \mathrm{exp}(-|y_k-h(\pmb{x}_{k}^{ip})|^2/\sigma^2),
\end{equation}
where 
\begin{equation}
\!\!	h(\pmb{x}_{k}^{ip})=\alpha \sqrt{p_{TX}}\mathbf{w}\left(\bar{\theta}\right)\bm{\alpha}_{RX}\left(\theta\right)\bm{\alpha}_{RIS}^H\left(\pmb{x}_{e,k}^{ip}, \pmb{x}_{a,k}^{ip}\right)\mathbf{e}_k.
\end{equation}
The normalized weights $\bar{\omega}_k^{ip}$ are computed as
\begin{equation}
	\bar{\omega}_k^{ip}=\omega_{k}^{ip}/\sum\nolimits_{ip=1}^{N_s}\omega_{k}^{ip}.
\end{equation}
The detailed algorithm for the channel tracking problem via our PF-WCA scheme is described in Algorithm \ref{SIRPF}.
\begin{algorithm}
	\caption{PF-WCA scheme}
	\label{SIRPF}
	\textbf{Initialization}: for $ip=1,2,\cdots,N_s$, generate $\pmb{x}_0^{ip} \sim p(\pmb{x}_0)$, and assign the weight $\omega_{0}^{ip}=1/N_s$.\\
	\textbf{Recursive update} 
	: \\
	\For{$k=1,2,\cdots,K$, where $K$ is the total number of time slots}{
		\For{$ip=1:N_s$}{
			-- Draw particles: $\pmb{x}_k^{ip}\sim q(\pmb{x}_k|\pmb{x}_{k-1}^{ip})$. \\
			-- Calculate weights: $\omega_{k}^{ip} = p(y_k|\pmb{x}_{k}^{ip})$.
		}
		Calculate the total weight: $t=\sum_{ip=1}^{N_s}\omega_{k}^{ip}$. \\
		\For{$ip=1:N_s$}{
			-- Normalize: $\bar{\omega}_{k}^{ip}=t^{-1}\omega_{k}^{ip}$.
		}
	    Estimate: $\hat{\pmb{x}}_k=\sum_{ip=1}^{N_s}\bar{\omega}_{k-1}^{ip}\pmb{x}_{k}^{ip}$. \\
	    Update phase shifts: Apply eq. \eqref{BA method}. \\
		Resample using Algorithm \ref{resampling}: \\
		$\left[\{\pmb{x}_k^{ip},\tilde{\omega}_{k}^{ip}\}_{ip=1}^{N_s}\right]=\mathrm{Resample}\left[\{\pmb{x}_{k}^{ip},\bar{\omega}_{k}^{ip}\}_{ip=1}^{N_s}\right]$.
	}
\end{algorithm}
In Algorithm \ref{SIRPF}, to improve the tracking performance, we update the phase shifts at the end of each iteration via the beam-matching (BA) method as
\begin{equation}
\mathbf{e}_{k+1}= \bm{\alpha}_{out}\left(\phi_{e},\phi_{a}\right)\odot \bm{\alpha}_{in}^{*}\left(\psi_{e,k},\psi_{a,k}\right),
\label{BA method}
\end{equation}
where the phase shifts at time slot $k$ are updated based on the hidden states tracked at time slot $k-1$.

In addition, we employ the resampling algorithm to mitigate the impairment caused by particle degeneracy. The resampling algorithm is given in Algorithm \ref{resampling}.
\begin{algorithm}
	\caption{Resampling}
	\label{resampling}
	\KwIn{$\{\pmb{x}_{k}^{i},\omega_{k}^{i}\}_{i=1}^{N_s}$} 
	\KwOut{$\{\pmb{x}_{k}^{j*},\omega_{k}^{j}\}_{j=1}^{N_s}$}
	$\bullet$ Initialize the CDF: $c_1=0$. \\
	\For{$i=2:N_s$}{
		-- Construct CDF: $c_i=c_{i-1}+\omega_{k}^{i}$.
	} 
	$\bullet$ Start at the bottom of the CDF: $i=1$. \\
	$\bullet$ Draw a starting point: $u_1\sim\mathcal{U}(0,N_s^{-1})$. \\
	\For{$j=1:N_s$}{
		-- Move along the CDF: $u_j=u_1+N_{s}^{-1}(j-1)$. \\
		\While{$u_j>c_i$}{
			$i=i+1$.
		}
		-- Assign sample: $\pmb{x}^{j*}_k=\pmb{x}^i_k$. \\
		-- Assign weight: $\omega_{k}^{j}=N_s^{-1}$. \\
	}
\end{algorithm}

\section{Simulation Results}
In this section, we assess the performance of our proposed PF-WCA scheme through the simulation. To demonstrate the benefits brought by the PF-WCA scheme, we choose a variant of EKF proposed in the recent paper \cite{RIStrackingCompared} as the benchmark, and the detailed derivations are shown in Appendix A.
Assume that the BS is located at $(60m,40m,40m)$ and the RIS is located at $(30m,0m,50m)$. The initial location of the user is at $(0m,20m,0m)$. The number of antennas at the BS is $N_r=16$. The carrier frequency is $f_c=28$ GHz.
We employ the normalized mean square error (NMSE) of $\mathbf{H}$ as the performance metric.
The NMSE is defined as $\frac{\mathbb{E}\{\|\hat{\mathbf{H}}_k-\mathbf{H}_k\|_F^2\}}{\mathbb{E}\{\|\mathbf{H}_k\|_F^2\}}$, where  $\mathbf{H}_k=\bm{\alpha}_{RX}\left(\theta\right)\bm{\alpha}_{RIS}^H\left(\phi_{e},\phi_{a},\psi_{e,k},\psi_{a,k}\right)\mathbf{e}_k$ is the true channel matrix and $\hat{\mathbf{H}}_k$ is the estimated channel matrix at time slot $k$.
We carry out the simulations with $\psi_s=0.5^{\circ}$ and $\psi_r=0.5^{\circ}$ under various numbers of particles and different settings of the RIS. Each block comprises 100 time slots, and we obtain results after running 1500 blocks for all results.

\subsection{Tracking Performance Comparison}
\begin{figure}
	\centering
	\includegraphics[width=0.75\textwidth]{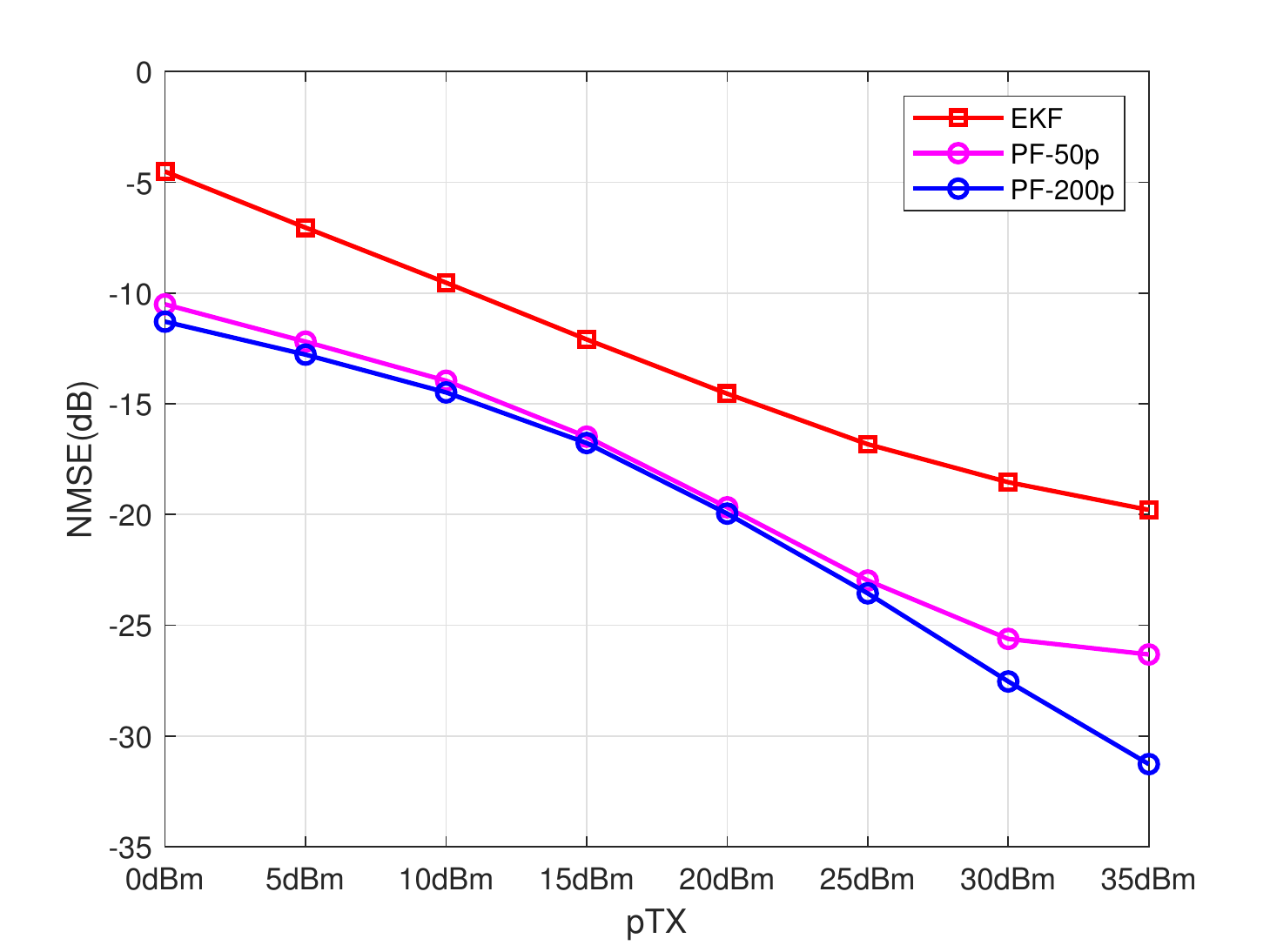}
	\caption{NMSE of $\mathbf{H}$ versus pTX between two schemes}
	\label{EKFvsPF}
\end{figure}
\rev{
In Fig. \ref{EKFvsPF}, we compare the tracking performance of these two tracking schemes. In this figure, the numbers of particles are set to 50 and 200 separately for the particle filter. The phase shifts matrix design method employed by both schemes is BA. As seen in Fig. \ref{EKFvsPF}, both EKF and PF schemes achieve a higher tracking accuracy with the increase of pTX. Moreover, PF has superior performance to EKF, and the superiority in tracking accuracy increases with the pTX. The reason is that the variation of the cascaded angles follows a non-Gaussian distribution. Hence, the EKF cannot parameterize the posterior density merely by the mean and covariance. On the other hand, PF characterizes the posterior function via the importance density, which shows its wide applications in various tracking tasks.
}

\subsection{The Impact of Particles}
\begin{figure}
	\centering
	\includegraphics[width=0.75\textwidth]{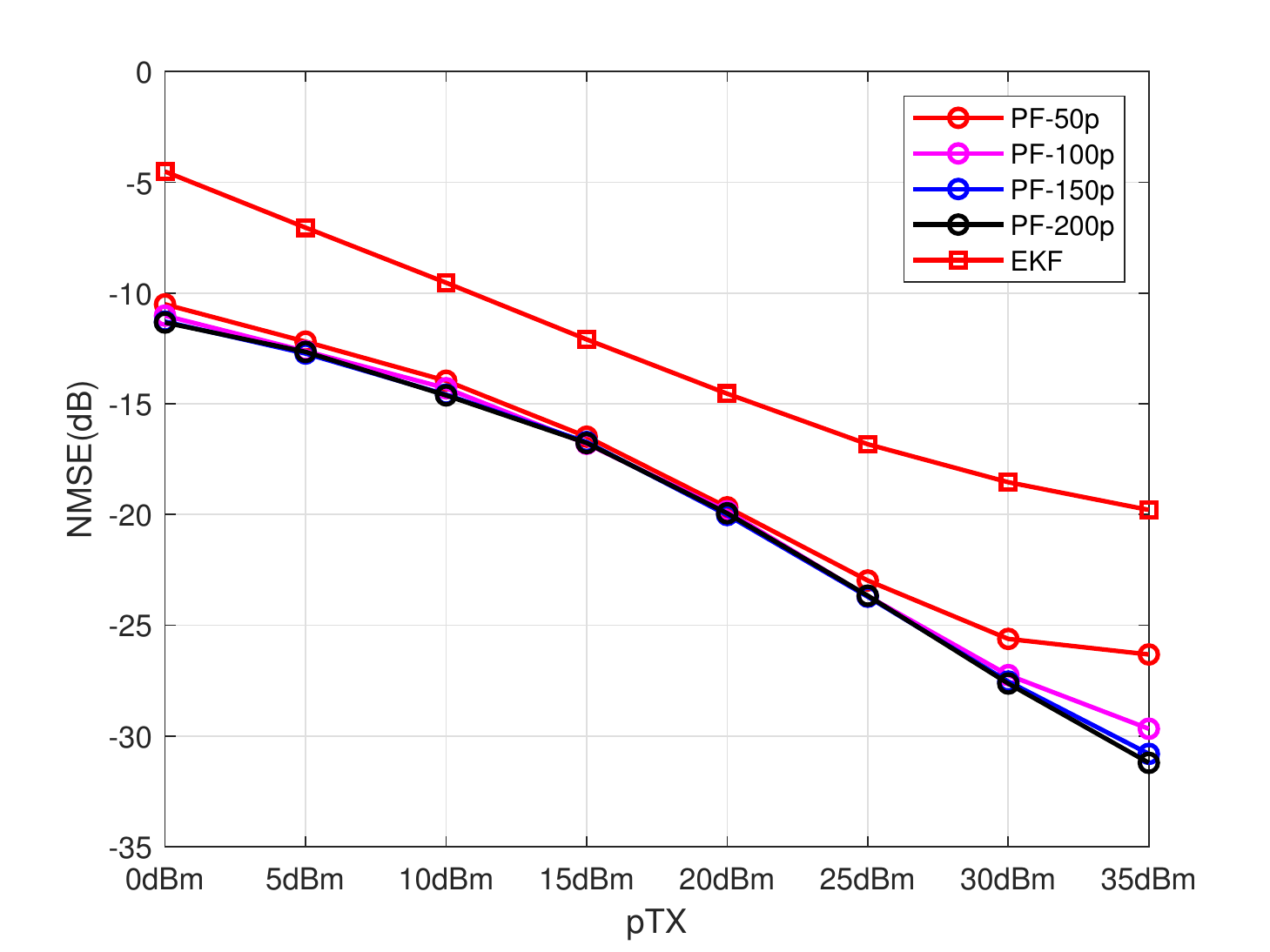}
	\caption{NMSE of $\mathbf{H}$ versus pTX via PF with various number of particles}
	\label{numParticles}
\end{figure}
\rev{
Fig. \ref{numParticles} illustrates the impact of the number of particles on the PF-WCA schemes. It is seen that the performance gap of the PF with different numbers of particles is relatively small with low pTX. Moreover, this gap becomes apparent with the increase of pTX. It is noted that the performance of PF with fewer particles levels off when the pTX reaches high due to particle impoverishment (PI). The resampling step leads to a loss of diversity among the particles as the resultant sample will contain many repeated particles. This PI problem is severe in the case of slight additive noise. In fact, in the case of very low additive noise, all particles will collapse to identical particles within a few slots, which degrades the tracking performance since identical particles cannot converge to the proper state. For particle filters, fewer particles may lead to PI more easily, which illustrates why the performance gap becomes enlarged with the pTX increases. 
It is noted that even with a small number of particles, the PF-WCA scheme still performs much better than the EKF scheme.
}

\subsection{The Impact of the RIS}
\begin{figure}
	\centering
	\includegraphics[width=0.75\textwidth]{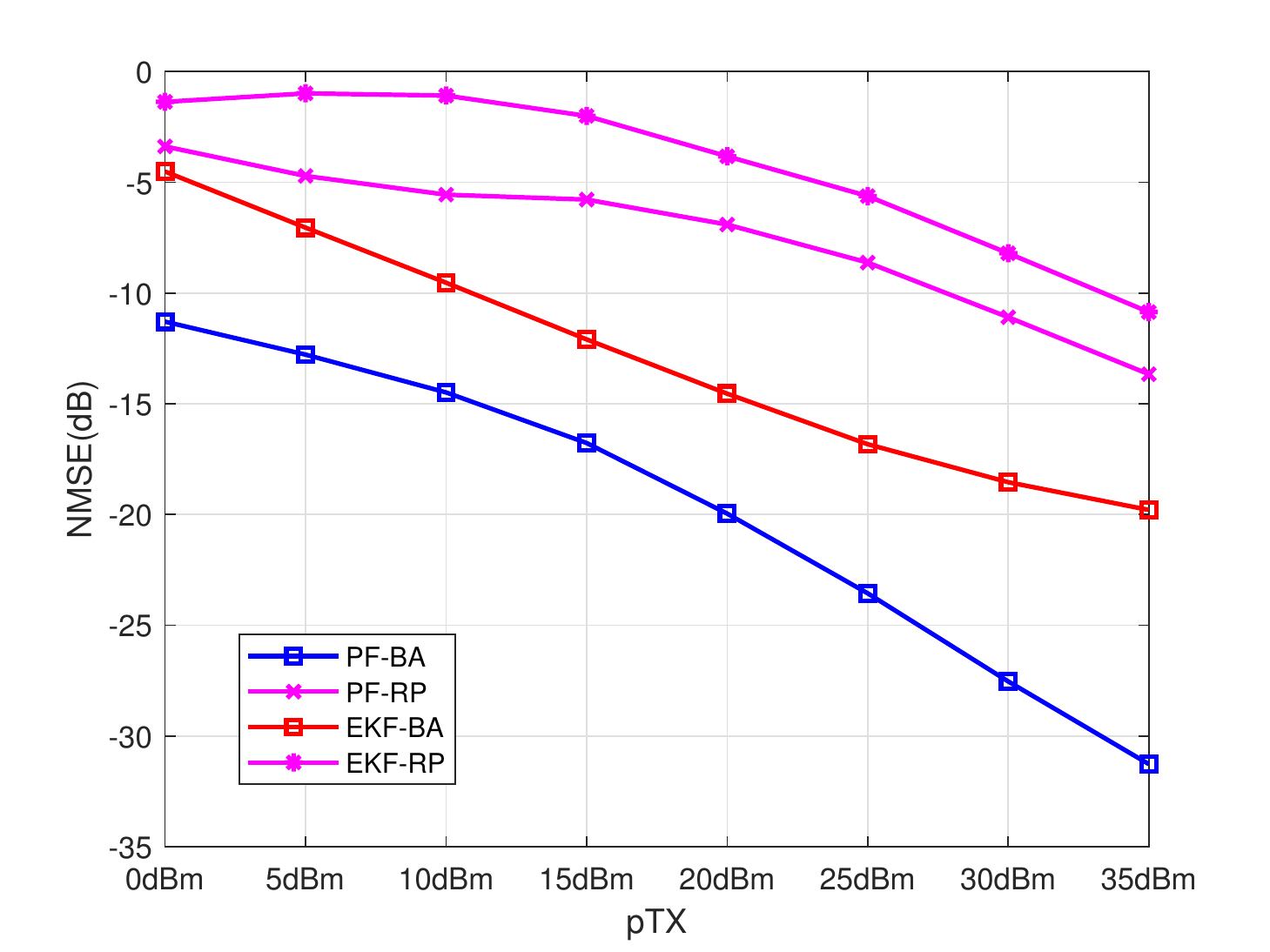}
	\caption{NMSE of $\mathbf{H}$ versus pTX with the different size and phase shifts of the RIS}
	\label{phaseshifts}
\end{figure}
\rev{
Fig. \ref{phaseshifts} shows the impact of phase shifts design of the RIS on the tracking accuracy. It is observed that the beam-matching approach has superior performance to the random phases for the both PF-WCA scheme and EKF scheme, which indicates that the tracking performance can be improved significantly via adjusting phase shifts properly. It is due to that proper phase shifts design can align the beam to enhance the signal strength, which improves the tracking performance significantly. In contrast, inappropriate phase shifts design may distort the received signal, and thus degrade the tracking accuracy. 
}

\section{Conclusion}
In this paper, we have investigated the RIS-aided mmWave channel tracking problem. A dual timescale variation model was adopted to characterize the channel, and we proposed a novel PF-WCA scheme to track the hidden states of the channel. Our proposed scheme showed its significantly improved performance over the recently proposed EKF-based scheme under a more general setting. In addition, we analyzed the the impact of the system parameters on the tracking performance.

\begin{appendices}
	\section{}
	To employ the EKF algorithm, Eq. \eqref{channel model1} is rewritten through the first-order Taylor expansion as 
	\begin{equation}
		\begin{aligned}
			y_k &=h(\pmb{x}_k)+z_k \\
			&=h\left(\pmb{x}_{k}-\hat{\pmb{x}}_{k|k-1}+\hat{\pmb{x}}_{k|k-1}\right)+z_k \\
			&\approx h\left(\hat{\pmb{x}}_{k|k-1}\right)+\frac{\partial h\left(\pmb{x}_{k}\right)}{\partial \pmb{x}_{k}}|_{\pmb{x}_{k}=\hat{\pmb{x}}_{k|k-1}}
			\left(\pmb{x}_{k}-\hat{\pmb{x}}_{k|k-1}\right)+z_k \\
			&=\mathbf{C}_{k}\left(\pmb{x}_{k}-\hat{\pmb{x}}_{k|k-1}\right)+z_k+d_k ,
		\end{aligned}
	\end{equation}
where $\hat{\pmb{x}}_{k|k-1}$ is the linear MMSE estimator of $\pmb{x}_k$ based on the observation $\{y_1, y_2,\cdots,y_{k-1}\}$, $\mathbf{C}_k=\frac{\partial h\left(\pmb{x}_{k}\right)}{\partial \pmb{x}_{k}}|_{\pmb{x}_{k}=\hat{\pmb{x}}_{k|k-1}}$ is the partial derivative of $h(\pmb{x})$ with respect to $\pmb{x}_{k}$, and $d_k=h\left(\hat{\pmb{x}}_{k|k-1}\right)-\mathbf{C}_{k}\hat{\pmb{x}}_{k|k-1}$. Then with the channel variation model, we can employ the Kalman filtering. In particular, $\pmb{x}_{k|i}$ is the linear MMSE estimator of $\pmb{x}_{k}$ based on the observation $\{y_1,\cdots,y_{i}\}$ , and $\mathbf{M}_{k|k-1}=\mathbb{E}\left[\left(\pmb{x}_k-\hat{\pmb{x}}_{k|k-1}\right)\left(\pmb{x}_k-\hat{\pmb{x}}_{k|k-1}\right)^H\right]$ is the minimum prediction mean square error (MSE) matrix, and $\mathbf{M}_{k|k}=\mathbb{E}\left[\left(\pmb{x}_k-\hat{\pmb{x}}_{k|k}\right)\left(\pmb{x}_k-\hat{\pmb{x}}_{k|k}\right)^H\right]$ is the MMSE matrix. 
In addition, $Q_v=\sigma^2$ and $\mathbf{Q}_u=\mathrm{diag}\left(\psi_s,\psi_r\right)$ are additive noise and covariance matrix of variation, respectively.
	\begin{algorithm}
		\caption{EKF}
		\textbf{Initialization}: $\hat{\pmb{x}}_{0|0}=\pmb{x}_{0}$, $M_{0|0}=\mathbf{0}$ \\
		\While{time slot $k$ still in the same block}{
			Predict: $\hat{\pmb{x}}_{k|k-1}=\hat{\pmb{x}}_{k-1|k-1}$ \\
			Calculate Minimum Prediction MSE: $\mathbf{M}_{k|k-1} = \mathbf{M}_{k-1|k-1}+\mathbf{Q}_u$ \\
			Calculate Kalman Gain Matrix: $\mathbf{K}_k = \mathbf{M}_{k|k-1}\mathbf{C}_k^H(Q_v+\mathbf{C}_k\mathbf{M}_{k|k-1}\mathbf{C}_k^H)^{-1}$ \\
			Correct: $\hat{\pmb{x}}_{k|k} = \hat{\pmb{x}}_{k|k-1}+\mathbf{K}_k\left(y_k-h\left(\hat{\pmb{x}}_{k|k-1}\right)\right)$ \\
			Estimate: $\hat{\pmb{x}}_{k|k}$ is the estimation of $\pmb{x}_k$ \\
			Calculte MMSE: $\mathbf{M}_{k|k} = (\mathbf{I}-\mathbf{K}_k\mathbf{C}_k)\mathbf{M}_{k|k-1}$ \\
		}
	\end{algorithm}

\revp{
\section{}
The filtered estimate is given by
\begin{equation}
	\hat{x}_t=\mathbb{E}\left[x_t|Y_{1:t}\right] .
\end{equation}
The mean of the conditional distribution is defined as
\begin{equation}
	\pi_{t|t}\left(dx_t\right)=P\left(X_t\in dx_t|Y_{1:t}=y_{1:t}\right) .
\end{equation}
The particle filters provide an estimate of these two quantities based on $N$ particles which we denote by $\hat{x}_t^N$ and $\pi_{t|t}^N\left(dx_t\right)$. For given $y_{1:t}$, $\hat{x}_t$ is a given scalar or vector, and $\pi_{t|t}\left(dx_t\right)$ is a given function. However, $\hat{x}_t^N$ and $\pi_{t|t}^N\left(dx_t\right)$ are random, since they depend on the randomly generated particles. We consider this scenario with a given $t$ and given observed outputs $y_{1:t}$. Hence, all stochastic quantifiers (like $\mathbb{E}$ and $w.p.1$) will be with respect to the random variables related to the particles. \\
First, we should introduce an additional notation. Given a measure $\nu$, a function $\phi$, and a Markov transition kernel $K$, and we denote the operations as
\begin{equation}
	\left(\nu,\phi\right)\triangleq\int\phi\left(x\right)\nu\left(dx\right)
	\label{initial}
\end{equation}
and
\begin{equation}
	K\phi\left(x\right)=\int K\left(dz|x\right)\phi\left(z\right).
\end{equation}
Hence, for any function $\phi:\mathbb{R}^{n_x}\leftarrow\mathbb{R}$, we have
\begin{equation}
	\mathbb{E}\left(\phi\left(x_t\right)|y_{1:t}\right)=\left(\pi_{t|t},\phi\right).
\end{equation}
It can be turned into the following recursive form,
\begin{align}
	\left(\pi_{t|t-1},\phi\right) &= \left(\pi_{t-1|t-1},K\phi\right)\\
	\left(\pi_{t|t},\phi\right) &= \frac{\left(\pi_{t|t-1},\phi\rho\right)}{\left(\pi_{t|t-1},\rho\right)}, \label{end}
\end{align}
where, $\rho$ is a density function with respect to a Lebesgue measure.
From Eq. \eqref{initial} to Eq. \eqref{end}, we have established a general optimal filter. \\
To prove the convergence of the particle filter, it is to prove
\begin{equation}
	\lim_{N\rightarrow\infty} \int\phi\left(x_t\right)\pi_{t|t}^N\left(dx_t\right)\rightarrow\mathbb{E}\left[\phi\left(x_t\right)|y_{1:t}\right]	
\end{equation}
Note that the $i$-th component of the estimate $\hat{x}_t^N$ is obtained for $\phi\left(x\right)=x[i]$ 
where \\ $x=\left[x[1],x[2],\cdots,x[n_x]\right]^T, i=1,2,\cdots,n_x$.  \\
In \cite{convergence}, the authors constructed a constant $C_\phi$ that depends on the function $\phi$, and find a bound on the fourth moment of the estimated conditional mean 
\begin{equation}
	\mathbb{E}\left[\left\|\int\phi\left(x_t\right)\pi_{t|t}^N\left(dx_t\right)-\int\phi\left(x_t\right)\pi_{t|t}\left(dx_t\right)\right\|^p\right]\le \frac{C_\phi}{N^2}.
\end{equation}
With $N$ increases to infinity, we have
\begin{equation}
	\lim_{N\rightarrow\infty} \mathbb{E}\left[\left\|\int\phi\left(x_t\right)\pi_{t|t}^N\left(dx_t\right)-\int\phi\left(x_t\right)\pi_{t|t}\left(dx_t\right)\right\|^4\right] \rightarrow 0,
\end{equation}
which means the estimated conditional mean is $L^4$ convergent. Hence, for any unbounded function $\phi$, we have that $\int\phi\left(x_t\right)\pi_{t|t}^N\left(dx_t\right)$ converges to $\int\phi\left(x_t\right)\pi_{t|t}\left(dx_t\right)$ in $L^4$ as $N\rightarrow\infty$. It is noted that the $i$-th component of the estimate $\hat{x}_t^N$ is obtained for $\phi\left(x\right)=x[i]$ where $x=\left[x[1],x[2],\cdots,x[n_x]\right]^T, i=1,2,\cdots,n_x$. Hence, we obtain 
\begin{equation}
	\lim_{N\rightarrow\infty}\hat{x}_t^N\rightarrow \hat{x}_t \quad w.p.1.
\end{equation}
So the particle filter state estimate will converge to the true estimate as the number of particles tends to infinity.
}
\end{appendices}

\bibliographystyle{IEEEtran}
\bibliography{myre}

\end{document}